\documentclass[%
 reprint,
 superscriptaddress,
 amsmath,amssymb,
 aps,
 prl,
]{revtex4-2}

\usepackage{siunitx}
\usepackage{color}
\usepackage{hhline}
\usepackage{mathrsfs}
\usepackage{graphicx}
\usepackage{adjustbox}
\usepackage{dcolumn}
\usepackage{bm}% bold math
\usepackage{multirow}
\usepackage{booktabs}
\usepackage{afterpage}
\usepackage{amsmath}
\usepackage{ulem}
\usepackage{physics}

\usepackage{ulem}

\usepackage{here} % force figures Here.

\usepackage[inline]{asymptote}

\usepackage[compat=1.1.0]{tikz-feynhand} % feynman diagram
\usepackage[version=4]{mhchem}

\usepackage{tikz, pgf, pgfplots, pgfplotstable}
\usepackage{tikz-3dplot}
\usetikzlibrary{math,calc}
\pgfplotsset{compat = newest}
\usepgfplotslibrary{groupplots} % LATEX and plain TEX
\pgfplotsset{compat = newest}

\pgfplotsset{
   table/search path={figures},
}

\graphicspath{{figures}}

\usepackage{multirow}

\usepackage[subpreambles=true,sort=true]{standalone}

\usepackage{comment}

\usepackage[colorlinks=true,citecolor=blue,linkcolor=blue,urlcolor=blue]{hyperref}
\usepackage{cleveref}
\crefname{equation}{Eq.}{Eq.}% {環境名}{単数形}{複数形} \crefで引くときの表示
\crefname{figure}{Fig.}{Fig.}% {環境名}{単数形}{複数形} \crefで引くときの表示
\crefname{table}{Table}{Table}% {環境名}{単数形}{複数形} \crefで引くときの表示
\crefname{section}{Sec.}{Sec.}% {環境名}{単数形}{複数形} \crefで引くときの表示
\crefname{appendix}{Appendix}{Appendix}% {環境名}{単数形}{複数形} \Crefで引くときの表示

\Crefname{equation}{Equation}{Equation}% {環境名}{単数形}{複数形} \Crefで引くときの表示
\Crefname{figure}{Figure}{Figure}% {環境名}{単数形}{複数形} \Crefで引くときの表示
\Crefname{table}{Table}{Table}% {環境名}{単数形}{複数形} \Crefで引くときの表示
\Crefname{section}{Section}{Section}% {環境名}{単数形}{複数形} \Crefで引くときの表示
\Crefname{appendix}{Appendix}{Appendix}% {環境名}{単数形}{複数形} \Crefで引くときの表示

\usepackage{threeparttable} %https://qiita.com/kumamupooh/items/38795811fc6b934a950d
\pgfplotsset{
   table/search path={figures},
}

\graphicspath{{figures}}

 \makeatletter
 \providecommand*{\input@path}{}
 \g@addto@macro\input@path{{docs/}{include/}}% append
 \makeatother
 
\begin{document}

\title{Data-Assimilated Crystal Growth Simulation for Multiple Crystalline Phases}

\author{Yuuki Kubo}
 \email{yuuki.kubo@phys.s.u-tokyo.ac.jp}
\affiliation{Department of Physics, The University of Tokyo, 7-3-1 Hongo, Bunkyo-ku, Tokyo 113-0033, Japan}
\author{Ryuhei Sato}
\affiliation{Department of Materials Engineering, The University of Tokyo, 7-3-1 Hongo, Bunkyo-ku, Tokyo 113-8656, Japan}
\author{Yuansheng Zhao}
\affiliation{Institute of Materials and Systems for Sustainability, Nagoya University, Furo-cho, Chikusa-ku, Nagoya, Aichi 464-8601, Japan}
\author{Takahiro Ishikawa}
\author{Shinji Tsuneyuki}
\affiliation{Department of Physics, The University of Tokyo, 7-3-1 Hongo, Bunkyo-ku, Tokyo 113-0033, Japan}

\date{\today}

\begin{abstract}
To determine crystal structures from an X-ray diffraction (XRD) pattern containing multiple unknown phases, a data-assimilated crystal growth (DACG) simulation method has been developed.
The XRD penalty function selectively stabilizes the structures in the experimental data, promoting their grain growth during simulated annealing.
Since the XRD pattern is calculated as the Fourier transform of the pair distribution function, the DACG simulation can be performed without prior determination of the lattice parameters.
We applied it to \ce{C} (graphite and diamond) and \ce{SiO2} (low-quartz and low-cristobalite) systems, demonstrating that the DACG simulation successfully reproduced multiple crystal structures.
\end{abstract}

\maketitle

% % intro
% \section{Introduction}\label{sec:intro}
Crystal structure determination is an essential process in materials science to predict and understand the properties of materials.
Experimental researchers mainly rely on X-ray diffraction (XRD) measurements or neutron diffraction (ND) spectroscopy to determine crystal structures.
To analyze new structures from powder XRD data, we first determine their lattice parameters.
Recent developments in crystallographic methodology have made it possible to systematically determine lattice parameters from single-phase XRD patterns~\cite{Visser:a07076,Werner:a25264,Altomare:ce5059,Boultif:ks0218,lebail_2004,Oishi-Tomiyasu:fs5064}.
On the other hand, because the peak assignment of XRD data significantly increases the number of wrong lattice parameter candidates, it is still challenging to find the correct lattice parameters from XRD data reflecting multiple unknown phases.

Computational science has also contributed to crystal structure determination, especially when it is difficult to determine the structure through experimental techniques~\cite{Woodley2008}.
In computational science, crystal structure has been predicted by finding the global minimum of the physical interatomic potential energy.
% Local optimization methods such as steepest descent and conjugate gradient methods are not suitable for global optimization on complex energy landscapes.
% This is because optimizations are easily captured by local minima near the initial structure when using these methods.
To overcome the potential energy barrier between local minima and the global minimum, metaheuristics methods based on random sampling~\cite{Pickard_2011}, particle-swarm optimization~\cite{PhysRevB.82.094116,WANG20122063}, genetic algorithms~\cite{10.1063/1.2210932,glass_uspexevolutionary_2006,PhysRevB.101.214106}, \textit{etc.}, have been employed with first-principles calculations in recent years.
Indeed, these methods successfully determined the new structures such as high-temperature superconducting hydrides under ultra-high pressure~\cite{10.1063/1.4874158,Duan2014,Drozdov2015,Einaga2016,PhysRevLett.119.107001,10.1073/pnas.1704505114,10.1002/anie.201709970,PhysRevLett.122.027001,Drozdov2019}.
In addition, combined with machine learning potentials, they can correspond to the variable composition analysis~\cite{PhysRevB.99.064114,PhysRevB.106.014102,PhysRevB.109.094106}.
Nevertheless, since these methods still require high computational costs, there is also a demand for computational approaches directly referring to the experimental data.
% In addition, there is a problem that there is no guarantee that the structures obtained by these methods coincide with those experimentally synthesized.

Under such circumstances, one possible way to accomplish rapid and accurate optimization is to supplement theoretical simulations with experimental data~\cite{Deem1992,10.1063/1.477812,Putz:zm0055,LANNING2000296,Coelho:ks0007}, and we have developed an experimental data-assimilated molecular dynamics (DAMD) simulation using XRD and ND data.
% To utilize experimental XRD and neutron scattering data, we have introduced a penalty function based on the dissimilarity between the XRD data calculated from the structures in the simulation and the experimental XRD data.
Our previous study confirmed that the DAMD simulation can efficiently determine the target crystalline~\cite{PhysRevMaterials.2.053801,10.1063/5.0125553} and amorphous~\cite{ZHAO2023122028} structures.
% In our previous studies, we have confirmed that XRD data-assimilated molecular dynamics (DAMD) simulations introducing this penalty function enables to efficiently determine the target crystalline~\cite{PhysRevMaterials.2.053801,10.1063/5.0125553} and amorphous~\cite{ZHAO2023122028} structures.
However, these simulations require the lattice parameters to be determined in advance~\cite{10.1063/5.0125553} and do not work well when XRD data contain peaks from multiple unknown crystalline phases.
In this study, we propose a data-assimilated crystal growth (DACG) simulation method, which is applicable to the structure determination even in those cases.
% This new scheme allows crystal structure determination not only when the lattice parameters are not obtained in advance, but also when XRD data contain peaks from multiple unknown crystalline phases.

\begin{figure}
\includegraphics{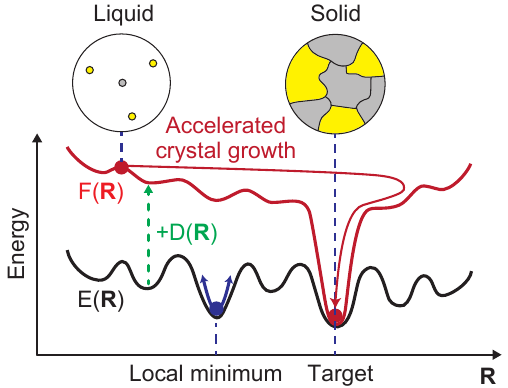}
\caption{\label{fig:met}Schematic image of the data-assimilated crystal growth (DACG) simulation.
The DACG simulation employs a simulation cell that is sufficiently large compared to the unit cells of the crystal structures.
Crystal growth simulations are accelerated by changing the potential energy surface from the interatomic potential $E(\vb{R})$ to the cost function $F(\vb{R})$, where $\vb{R}$ represents the atomic coordinates.
Note that the cost function $F(\vb{R})$ is the linear sum of the interatomic potential $E(\vb{R})$ and the XRD penalty function $D(\vb{R})$.}
\end{figure}

% theory
% \section{Methods}\label{sec:methods}
Figure~\ref{fig:met} shows a schematic image of the DACG simulation proposed in this study.
Here, we assume that the experimental XRD data contain peaks from multiple unknown phases, and the lattice parameters cannot be readily determined.
Therefore, in the DACG simulation, we employ large simulation cells including a few thoudsand atoms to obtain grains of the target crystal structures rather than the perfect crystals. 
% The idea of LDAMD consists of the following two components: (i) crystal growth simulation based on simulated annealing (SA) is performed using simulation cells that are sufficiently large compared to the unit cells to determine crystal structures without prior determination of the lattice parameters, and (ii) the cost function $F$ described below is used instead of the interatomic potential $E$ during simulated annealing to accelerate the crystal growth simulation.

To obtain crystal grains, MD simulations of crystal growth are performed and accelerated by employing the cost function $F$~\cite{PhysRevMaterials.2.053801,10.1063/5.0125553,ZHAO2023122028}:
\begin{align}
F(\vb{R}) = E(\vb{R}) + \alpha N D[I_\mathrm{ref}(Q), I_\mathrm{calc}(Q; \vb{R})] \label{eq:one},
\end{align}
where $\vb{R}$ is the atomic coordinates and $E$ is the interatomic potential energy.
$I_\mathrm{ref}$ and $I_\mathrm{calc}$ are the XRD intensity referred from experimental data and that calculated from a structure in the simulation, respectively.
$D$ is the XRD penalty function that represents the dissimilarity between $I_\mathrm{ref}$ and $I_\mathrm{calc}$.
$\alpha$ is the weight parameter, $N$ is the number of atoms, and $Q$ is the magnitude of the scattering vector.
Most of the structures inconsistent with the experimental data are destabilized by the cost function $F$ as shown in Fig.~\ref{fig:met}.
Therefore, DA can increase the probability of reaching the target structures through optimization.

Following our previous study~\cite{10.1063/5.0125553}, we used a penalty function $D$ based on the correlation coefficient:
\begin{align}
D = 1 - \frac{\int\dd{Q}(I_\mathrm{ref} - \overline{I}_\mathrm{ref})(I_\mathrm{calc} - \overline{I}_\mathrm{calc})}{\sqrt{\int\dd{Q}(I_\mathrm{ref} - \overline{I}_\mathrm{ref})^2}\sqrt{\int\dd{Q}(I_\mathrm{calc} - \overline{I}_\mathrm{calc})^2}} \label{eq:two},
\end{align}
where $\overline{I}$ is the averaged XRD intensity over the integral range of $Q$.
By defining $D$ in this way, we can simultaneously compare the peak positions and intensities of $I_\mathrm{ref}$ and $I_\mathrm{calc}$.
% In our previous study~\cite{10.1063/5.0125553}, a penalty function $D$ based on the correlation coefficient was used instead.
% However, when simulation cells are large, as in this study, using a penalty function $D$ based on the correlation coefficient increases the probability that the optimization will fall into irrelevant structures.
% This is because it is possible that $\overline{I}_\mathrm{calc}$ cannot be neglected in comparison to $I_\mathrm{calc}$, even though $I_\mathrm{calc} - \overline{I}_\mathrm{calc}$ itself is roughly proportional to $I_\mathrm{ref} - \overline{I}_\mathrm{ref}$.
% The above problem can be avoided by using a penalty function $D$ based on cosine similarity, where $\overline{I}_\mathrm{calc}$ is not subtracted from $I_\mathrm{calc}$.

In addition, we adopted the following formula to calculate $I_\mathrm{calc}$:
\begin{align}
\frac{I_\mathrm{calc}}{L(2\theta)} &\propto \sum_{\mu}c_{\mu}f_{\mu}^2(Q) + \sum_{\mu, \nu}c_{\mu}c_{\nu}f_{\mu}(Q)f_{\nu}(Q) \nonumber\\
&\times\int_0^{r_\mathrm{cutoff}} \dd{r}4\pi r^2\rho_0\qty[g_{\mu\nu}(r) - 1]\frac{\sin Qr}{Qr} \label{eq:three},
\end{align}
where $2\theta$ is the angle between the incident and scattered X-rays, $L(2\theta)$ is the polarization factor, $c_{\mu}$ is the ratio of atomic species $\mu$, $f_{\mu}(Q)$ is the atomic form factor, $\rho_0$ is the number of atoms per volume, and $g_{\mu\nu}(r)$ is the partial pair distribution function (PDF) between atomic species $\mu$ and $\nu$.
% Practically, we introduce a cutoff distance $r_\mathrm{cutoff}$ as an upper bound on the interval of integration in Eq.~(\ref{eq:three}).
$r_\mathrm{cutoff}$ is the cutoff radius for calculating PDF.
Since Eq.~(\ref{eq:three}) does not include any lattice parameters and Laue indices directly, we can perform the DACG simulation without any prior determination of the lattice parameters.
Furthermore, this method reduces the computational cost by making the cutoff distance $r_\mathrm{cutoff}$ smaller.
% Therefore, although the peak resolution is decreased because the peak width of the calculated XRD pattern is inversely proportional to the cutoff, this method allows us to calculate XRD patterns and the penalty function $D$ without the information on the lattice parameters of the target structures as indicated by the absence of Laue indices in the Eq.~(\ref{eq:three}).
On the other hand, the drawback of this method is that the resolution of the calculated XRD peak position becomes inversely proportional to $r_\mathrm{cutoff}$.
% This means that the target structure can be determined by this method without prior determination of the lattice parameters.

When the DACG simulations are performed using multi-phase XRD data as the reference, a polycrystalline structure containing multiple different crystal structures may be obtained in a single simulation cell.
To extract the grains of each phase in those cases, we perform K-means clustering~\cite{scikit-learn} based on the site-specific radial distribution function (SSRDF) around each atom $i$ defined as
\begin{align}
g_i(r) = \frac{1}{4\pi r^2\rho_0}\sum_{j\neq i}\delta(r-r_{ij})f_\mathrm{cut}(r),
\end{align}
where $\rho_0$ is the number of atoms per volume, and $r_{ij}$ is the distance between atoms $i$ and $j$.
Here, $f_\mathrm{cut}(r)$ is a smooth cutoff function defined as
\begin{align}
f_\mathrm{cut}(r) = \left\{ \,
    \begin{aligned}
        &\frac{1}{2}\qty[\cos\qty(\pi\frac{r}{R})+1] &(r < R),\\
        &0 &(r \geq R),
    \end{aligned}
\right.
\end{align}
using cutoff radius $R$.

% result
% \section{Results and Discussion}

% \subsection{Computational details}\label{sec:details}
To test the effectiveness of the DACG simulation, we applied it to the structure determination of \ce{C} (graphite and diamond) and \ce{SiO2} (low-quartz and low-cristobalite) polymorphs.
We used the Large-scale Atomic/Molecular Massively Parallel Simulator (LAMMPS)~\cite{PLIMPTON19951,THOMPSON2022108171} package and the external code for calculating the penalty function implemented in our previous study~\cite{ZHAO2023122028}.
The long-range carbon bond order potential (LCBOP)~\cite{PhysRevB.68.024107} for \ce{C} and the Tsuneyuki potential~\cite{PhysRevLett.61.869} for \ce{SiO2} have been employed for the physical interatomic potential energy.
The temperature was controlled by the velocity scaling method~\cite{WOODCOCK1971257}, and the time step was set to $\SI{0.5}{}$ and $\SI{1}{\femto\second}$ for \ce{C} and \ce{SiO2} simulations, respectively.
XRD patterns from $\SI{0}{}$ to $\SI{4}{\angstrom}{}^{-1}$ were used to calculate the penalty function $D$.
For calculating $I_\mathrm{calc}$ from structures during the DACG simulation, we set $r_\mathrm{cutoff}$ to $\SI{15}{}$ and $\SI{21}{}$-$\SI{25}{\angstrom}$ for \ce{C} and \ce{SiO2} cases, respectively.
As initial atomic configurations, $\SI{6000}{}$ and $\SI{10200}{}$ atoms were randomly placed in cubic boxes for \ce{C} and \ce{SiO2} cases, respectively.
Here, the simulation cell size was determined so that the atomic density was consistent with the density of each desired crystal structure~\cite{FAYOS1999278,Levien,+1973+274+298}.
We performed simulated annealing with the cost function $F$ (DA, data assimilation) and that with the interatomic potential $E$ (SA, normal simulated annealing).
First, since the initial configurations were energetically too unstable, we performed short MD simulations with small time steps without the XRD penalty function to relax the structure.
Next, the temperatures were set at $\SI{14000}{}$ and $\SI{10000}{\kelvin}$ for \ce{C} and \ce{SiO2}, respectively, and then decreased linearly to $\SI{0}{\kelvin}$ in $\SI{0.5}{}$ and $\SI{5}{\nano\second}$.
The weight parameter $\alpha$ of the XRD penalty function in Eq.~(\ref{eq:one}) was set to $\SI{10}{\electronvolt}$ for both \ce{C} and \ce{SiO2} cases in this study.

% \subsection{Application to single-phase materials}\label{sec:single}
\begin{figure*}
\includegraphics{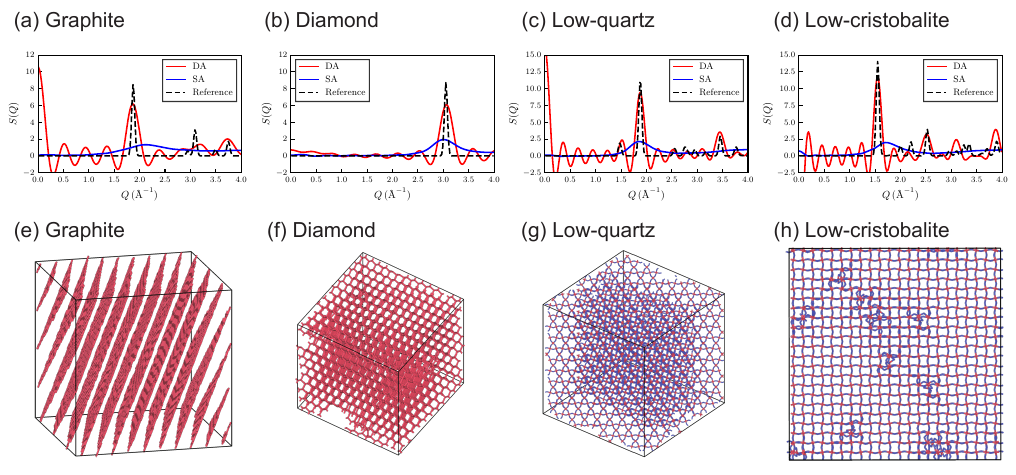}
\caption{\label{fig:res1}(Top) (a)-(d) Structure factors $S(Q)$ corresponding to the ideal structures (black dashed lines), and those of the structures obtained by normal simulated annealing (SA, blue solid lines) and data-assimilated simulations (DA, red solid lines) as functions of wavenumber $Q$. (Bottom) Snapshots of the structures obtained after the DACG simulations using single-phase XRD patterns of the ideal (e) graphite, (f) diamond, (g) low-quartz, and (h) low-cristobalite structures. These snapshots were drawn by OVITO~\cite{Stukowski_2010}.}
\end{figure*}
\begin{figure}
\includegraphics{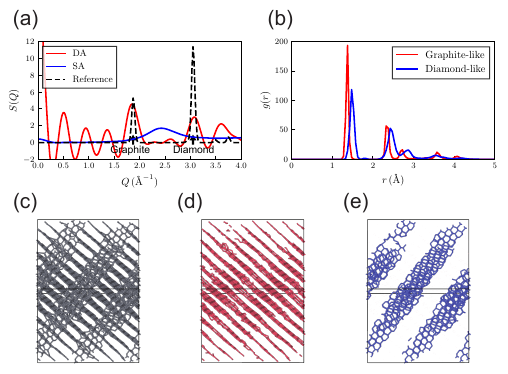}
\caption{\label{fig:res2}(a) Structure factor $S(Q)$ corresponding to the reference XRD data (black dashed line), the structure obtained by normal simulated annealing (SA, blue solid line), and the structure obtained by the DACG simulation (DA, red solid line). (b) SSRDF averaged over center atoms in each phase. (c) A snapshot of the structure obtained after the DACG simulation using XRD data containing peaks of both graphite and diamond as the reference. (d) and (e) Atoms in the clusters obtained by K-means clustering. }
\end{figure}

First, we performed simulated annealing with the cost function $F$ (Eq.~(\ref{eq:one})) using single-phase XRD data of graphite~(\ce{C}), diamond~(\ce{C}), low-quartz~(\ce{SiO2}) and low-cristobalite~(\ce{SiO2}).
Figures~\ref{fig:res1}(a)-(d) show the structure factors $S(Q)$, corresponding to the reference XRD pattern used for the DACG simulations (black dashed lines), and those calculated from the structures obtained by DA (red lines) and SA (blue lines). 
% The upper panels of Fig.~\ref{fig:res1} show the snapshots of the structures obtained after the DAMD simulations.
% The lower panels of Fig.~\ref{fig:res1} show the static structure factors $S(Q)$, corresponding to the reference XRD pattern used for the DAMD simulations, and the ones calculated from the structures obtained by DA and SA.
% As shown in the lower panels, no peaks appear in $S(Q)$ for the structures obtained by SA, suggesting that no ordered structures were obtained.
As shown in Figs.~\ref{fig:res1}(a)-(d), no Bragg peaks appear in $S(Q)$ for the structures obtained by SA, suggesting that no ordered structures were obtained.
% In contrast, DA reproduces the strongest peak in the reference $S(Q)$ for each of the four cases and the peaks in the higher angle region for graphite, low-quartz, and low-cristobalite.
On the other hand, peak positions of $S(Q)$ obtained by DA are comparable with those in the reference $S(Q)$.
This suggests that the target structures were successfully obtained by DA owing to the XRD penalty function $D$.
Figures~\ref{fig:res1}(e)-(h) show the snapshots of the structures obtained after the DACG simulations.
As shown in the snapshots, we successfully obtained the target structures: a layered structure like graphite in Fig.~\ref{fig:res1}(e), a three-dimensional network originated from $\mathrm{sp}^3$ hybrid orbitals formed over the whole simulation cell in Fig.~\ref{fig:res1}(f), a triple-helical structure with some disordered parts in Fig.~\ref{fig:res1}(g), and a network consisting only of six-membered rings in Fig.~\ref{fig:res1}(h).
These results suggest that the DACG simulations can determine the target structures correctly from single-phase XRD patterns without prior determination of the lattice parameters.
Note that the small oscillations appearing in $S(Q)$ in the DA cases are caused by the Fourier transformation of PDF with the finite cutoff distance $r_\mathrm{cutoff}$ in Eq.~(\ref{eq:three}).
Next, we conducted DACG simulations to determine multiple crystal structures from multi-phase reference XRD data.
We used $1$:$1$ mixture of the XRD patterns of ideal graphite and diamond structures as the reference XRD data (black dashed line) as shown in Fig.~\ref{fig:res2}(a).
The atomic density was set to the average of those of graphite and diamond.
Figure~\ref{fig:res2}(a) also shows the structure factors $S(Q)$ of the structures obtained by DA and SA.
As shown in Fig.~\ref{fig:res2}(a), peak positions of $S(Q)$ obtained after the DACG simulation (red solid line) agrees well with those of the ideal graphite and diamond peaks in the reference $S(Q)$, suggesting that the structure obtained by DA contains ordered structures related to the target ones.
On the other hand, any peaks in the reference $S(Q)$ were not reproduced by SA, suggesting that SA without the penalty function $D$ did not even yield any ordered structures.
We can also see that the structure obtained by the DACG simulation indeed has some ordered structures shown in Fig.~\ref{fig:res2}(c).
However, it is difficult to conclude from only the snapshot whether this structure contains both graphite and diamond structures simultaneously.
Therefore, we conducted K-means clustering to extract crystal grains from the obtained structure.
Here, the number of groups for clustering was set to two so that the atoms could be classified into graphite and diamond phases, and the cutoff distance for the SSRDF calculation was set to $\SI{5}{\angstrom}$.
As the result of the clustering, the atoms were classified into two phases as shown in Figs.~\ref{fig:res2}(d) and (e) which form graphite-like and diamond-like structures, respectively.
Furthermore, the SSRDFs averaged over the center atoms in Figs.~\ref{fig:res2}(d) (red line) and (e) (blue line) are shown in Fig.~\ref{fig:res2}(b).
As shown in this figure, it is confirmed that there are groups of atoms with different distances to the first and second nearest neighboring atoms and that the clustering successfully distinguished them.
Thus, we concluded that, in the case of multi-phase \ce{C}, graphite and diamond were obtained simultaneously by a single DACG simulation, and the grain of each phase was successfully extracted by K-means clustering using the SSRDF as an atomic fingerprint.
% Figure~\ref{fig:res2}(d) shows the static structure factors $S(Q)$, corresponding to the XRD pattern used as the reference for the DAMD simulations and the ones calculated from the structures obtained by DA and SA.
% None of the peaks in the reference $S(Q)$ were reproduced by SA, suggesting that SA failed to yield ordered structures.
% In contrast, the two strong peaks originated from the strongest peaks in graphite and diamond in the reference $S(Q)$ were successfully reproduced by DA, suggesting that the structure obtained by DA contains ordered structures related to the target ones.
% To analyze the obtained structure in more detail, we observed the non-averaged radial distribution function (RDF) around each atom in the structure and classified them according to the RDF using K-means clustering.

% \subsubsection{Multi-phase \ce{SiO2}}
In the case of multi-phase \ce{SiO2}, all the target structures (low-quartz and low-cristobalite) were successfully reproduced by repeating the DACG simulation multiple times.
\begin{figure}
\includegraphics{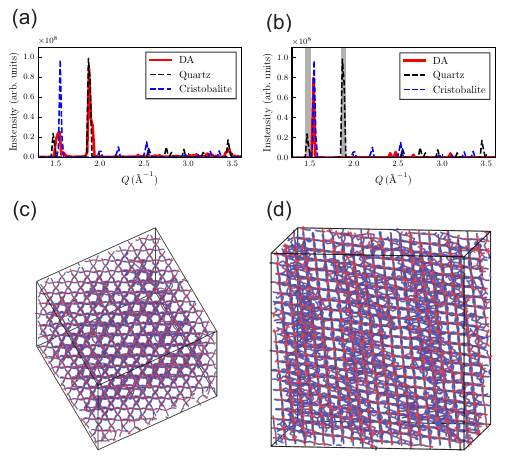}
\caption{\label{fig:res3}(a) XRD patterns of low-quartz (black dashed line), low-cristobalite (blue dashed line), and the structure obtained after the first DACG simulation (red solid line). (b) Reference XRD pattern and the one of the structure obtained after the second DACG simulation, where XRD patterns in the range of $1.45\leq Q\leq 1.5$ and $1.85\leq Q\leq 1.9$ (masked area) were not used during the simulation. Note that all XRD patterns in these figures were calculated by using the lattice parameters of the simulation cell and Laue indices to analyze the obtained structure in detail. (c) A snapshot of the structure obtained by the first DACG simulation, using XRD data containing peaks of both low-quartz and low-cristobalite as the reference. (d) A snapshot of the structure obtained after the second DACG simulation.}
\end{figure}
Figure~\ref{fig:res3}(a) shows the XRD patterns of ideal low-quartz and low-cristobalite, which were mixed in a $2.5$:$1$ ratio and used as the reference XRD data in the DACG simulations.
The mixing ratio of the XRD patterns was determined so that the maximum intensity of low-quartz and low-cristobalite XRD patterns take the same value.
The atomic density was set to the average of those of low-quartz and low-cristobalite.
Figure~\ref{fig:res3}(a) also shows the XRD pattern calculated from the structure obtained after the first DACG simulation.
This XRD pattern was calculated using the lattice parameters of the simulation cell and Laue indices to analyze the obtained structure in detail.
Note that during the DACG simulation, XRD patterns were still calculated using Eq.~(\ref{eq:three}).
As shown in the figure, the first and second peaks of quartz were reproduced by the first DACG simulation.
On the other hand, none of the peaks of cristobalite were reproduced by the first DACG simulation.
This suggests that the first DACG simulation yielded only quartz-like ordered structures.
Figure~\ref{fig:res3}(c) shows the snapshot of the structure obtained by the first DACG simulation, in which a quartz-like triple-helical structure exists.
% Note that the two peaks at $29^\circ$ and $31^\circ$ were not reproduced by the DAMD simulation, which may be due to the mixture of low- and high-cristobalite in the obtained structure.
% This suggests that it is difficult to distinguish such slight differences in structure through the DA method.

To obtain the crystal structures of the remaining phases, we conducted the DACG simulation again using the reference XRD data.
However, this time, the powder XRD pattern in the range highlighted in gray in Fig.~\ref{fig:res3}(b) was not used for calculating the XRD penalty function $D$ during the simulation, since the structure corresponding to the peaks in those ranges has been already obtained by the previous DACG simulation.
% Therefore, XRD patterns in the range of $20^\circ$ to $22^\circ$ near the strongest peak of low-cristobalite were not used to calculate the XRD penalty function $D$ when performing additional DAMD simulation.
Figure~\ref{fig:res3}(b) shows the XRD pattern calculated using the lattice parameters of the simulation cell and Laue indices from the structure obtained after the second DACG simulation.
As shown in this figure, unlike the first DACG simulation, the second one reproduced the first peak of cristobalite.
Furthermore, Fig.~\ref{fig:res3}(d) shows a snapshot of the structure obtained by the second DACG simulation, in which a cristobalite-like structure can be seen.
Thus, we conclude that in the case of multi-phase \ce{SiO2}, all the target structures were successfully obtained by the sequential DACG simulations, in which the XRD peaks related to the structures obtained at each DACG simulation are masked one by one.

% \section{Conclusions}\label{sec:conclusions}
In summary, we proposed a computer simulation method for efficiently growing and finding crystal structures by assimilating powder diffraction data from unknown crystalline phases.
This method will help determine the crystal structures of new materials in combinatorial synthesis or under extreme conditions.

\begin{acknowledgements}
This work was supported by JSPS KAKENHI Grant numbers JP18H05519, JP24K00544, and JP20H05644.
The computation in this work was partly performed by using the facilities of the Supercomputer Center, the Institute for Solid State Physics, The University of Tokyo.
Y.K. is supported by MEXT - Quantum Leap Flagship Program (MEXT Q-LEAP).
\end{acknowledgements}

% \input{appendix.tex}

% \newpage
% references 
\bibliographystyle{apsrev4-2}
%apsrev4-2.bst 2019-01-14 (MD) hand-edited version of apsrev4-1.bst
%Control: key (0)
%Control: author (72) initials jnrlst
%Control: editor formatted (1) identically to author
%Control: production of article title (-1) disabled
%Control: page (0) single
%Control: year (1) truncated
%Control: production of eprint (0) enabled
%	
% \bibliography{references/ref.bib}
\end{document}